\documentclass[conference]{IEEEtran}
\IEEEoverridecommandlockouts

\usepackage{cite}
\usepackage{color}
\usepackage{amsmath}

\usepackage{graphicx}
\usepackage{epsfig}
\usepackage{epstopdf}
\usepackage{times}
\usepackage{bm}
\usepackage{amssymb}
\usepackage{url}
\usepackage{flushend}
\usepackage{multirow}
\usepackage{array}
\usepackage{ntheorem}
\usepackage{algorithm}
\usepackage{algorithmic}        %用到的宏包，要自己改下
\usepackage{multirow}
\usepackage{graphics} % for pdf, bitmapped graphics files
\usepackage{subfigure}

\ifCLASSINFOpdf
\else
\fi
\usepackage{amsmath}
\usepackage{graphicx}
\usepackage{enumitem}

\begin{document}

\renewcommand{\IEEEQED}{\IEEEQEDopen}

\newtheorem{theorem}{Theorem}[section]
\newtheorem{lemma}[theorem]{Lemma}
\newtheorem{proposition}[theorem]{Proposition}
\newtheorem{corollary}[theorem]{Corollary}
\newtheorem{assumption}[theorem]{Assumption}

\newenvironment{proof}[1][Proof]{\begin{trivlist}
\item[\hskip \labelsep {\bfseries #1}]}{\end{trivlist}}
\newenvironment{definition}[1][Definition]{\begin{trivlist}
\item[\hskip \labelsep {\bfseries #1}]}{\end{trivlist}}
\newenvironment{example}[1][Example]{\begin{trivlist}
\item[\hskip \labelsep {\bfseries #1}]}{\end{trivlist}}
\newenvironment{remark}[1][Remark]{\begin{trivlist}
\item[\hskip \labelsep {\bfseries #1}]}{\end{trivlist}}

\newcommand{\qed}{\nobreak \ifvmode \relax \else
      \ifdim\lastskip<1.5em \hskip-\lastskip
      \hskip1.5em plus0em minus0.5em \fi \nobreak
      \vrule height0.75em width0.5em depth0.25em\fi}

\title{Cooperative Curve Tracking in Two Dimensions Without Explicit Estimation of the Field Gradient}

\author{\IEEEauthorblockN{Sarthak Chatterjee, Wencen Wu \thanks{The research work is supported by NSF grant CNS-1446461.}}
\IEEEauthorblockA{Department of Electrical, Computer and Systems Engineering\\
Rensselaer Polytechnic Institute\\
Troy, NY 12180\\
chatts3, wuw8@rpi.edu}}

%\and
%\IEEEauthorblockN{Homer Simpson}
%\IEEEauthorblockA{Twentieth Century Fox\\
%\and
%\IEEEauthorblockN{James Kirk\\ and Montgomery Scott}
%\IEEEauthorblockA{Starfleet Academy\\
%San Francisco, California 96678--2391\\


%Telephone: (800) 555--1212, Fax: (888) 555--1212}


\maketitle


\begin{abstract}
We design a control law for two agents to successfully track a level curve in the plane without explicitly estimating the field gradient. The velocity of each agent is decomposed along two mutually perpendicular directions, and separate control laws are designed along each direction. We prove that the formation center will converge to the neighborhood of the level curve with the desired level value. The algorithm is tested on some test functions used in optimization problems in the presence of noise. Our results indicate that in spite of the control law being simple and gradient-free, we are able to successfully track noisy planar level curves fast and with a high degree of accuracy.
\end{abstract}


% \fi
\IEEEpeerreviewmaketitle

\section{Introduction}

Mobile sensor networks are increasingly being used for cooperative collection of information that ultimately is used for tracking physical characteristics of the environment. Monitoring mechanisms of sensor networks allow us to keep track of environmental changes for long periods of time with great reliability. Using mobile sensor networks allow us to tackle the environmental monitoring problem with lesser computational power, since installing a large number of static sensors may not always be feasible from the point of view of cost. Very few mobile sensor networks can be used extremely efficiently to explore large environmental landscapes, hence allowing remote and dynamic monitoring of environmental changes. Mobile sensor networks have been used in data dissemination and collection, sensor network platforms, and motion monitoring, to talk of a few areas, \cite{Sayyed2015}, \cite{Amundson2009}, \cite{Zhu2010}.

In recent years, a significant body of work has been devoted to the problem of exploration of environmental boundaries, \cite{Clark2005}, \cite{Clark2007}, \cite{Hsieh2005} and \cite{Joshi2009}. Specifically, the work in \cite{Zhang2007} focused on planar motion control in which particles are controlled to converge and travel along a closed curve while maintaining formation. The work in \cite{ZhangCDC2007}  focused on large-scale level curve tracking of environmental scalar fields by using four moving sensor platforms for the cooperative exploration of a noisy scalar field. As a natural extension, \cite{Wu2011} explores the problem of tracking level curves of three-dimensional fields. They use a cooperative Kalman filter to estimate the field value and gradient at the formation center and also use Taubin's algorithm \cite{Taubin} for estimation of principal curvatures and principal directions for the lines of curvature in the field.
%The problem of curve tracking has been considered in a lot of works in the literature. Zhang and Leonard in \cite{Zhang2007} develop a method for motion control for particles on a plane. Starting from their initial positions within a planar compact set, the particles are controlled to converge and travel along a closed curve. The pairwise-distance between particles is also controlled to converge to constant values so as to ensure that the particles track the closed curve in unison. Zhang, Fiorelli and Leonard have also considered the problem of level-curve tracking in large-scale environmental scalar fields in \cite{ZhangCDC2007}. In their approach, four moving sensor platforms are used for the cooperative exploration of a noisy scalar field in the plane. A Kalman filter provides estimates of the field value, the gradient and the Hessian and the developed motion control law allows the formation center to move along the level curves of the field. Wu and Zhang in \cite{Wu2011} extend these techniques to successfully explore level surfaces of three-dimensional scalar fields. Their approach uses a cooperative Kalman filter which analyzes the agent sensor readings and estimates the value of the field and its gradient at the formation center. The steering control laws of the agents are so designed as to track curves on a level surface in the field. Another contribution of this work is that it uses Taubin's Algorithm \cite{Taubin} for successful estimation of principal curvatures and principal directions for the lines of curvature in the field.
\par Relevant work has also been carried out in the area of tracking planar level curves without explicit estimation of the field gradient. Information about the gradient is unavailable in most cases, and estimating the same is difficult because the latter requires the knowledge of field values at multiple locations. The major motivating factor for gradient-free tracking is the cost-effective scenario of a single or two mobile sensors with access to only instantaneous measurements of the field value. Works by Matveev \textit{et al}., \cite{Matveev2015}, \cite{Matveev2012}, use techniques in sliding mode control for gradient-free boundary tracking of an unknown dynamic scalar field using a planar mobile robot. More applications of gradient-free tracking can be found in \cite{Andersson2007}, \cite{Barat2003}, and \cite{Kemp2004}.
\par In this paper, we propose an algorithm that allows a two-agent system to track a level curve for a noisy field without explicitly computing the gradient of the field, while maintaining a fixed distance between the agents. We consider the problem of a two-agent system, both equipped with sensors capable of measuring the (noisy) level values of the scalar field in question. The velocities of the agents are decomposed along two mutually perpendicular directions, and the control laws are derived separately along both the directions. From the nonlinear dynamical system we obtain, we show that the two-agent system asymptotically converges to the desired level field value, and that the formation center can successfully track the level curve.
\par The motivation of this work is two-fold. We try to use as few agents as possible to achieve a fast rate of convergence to the level curve. As with any problem, there is a trade-off when we try to use less computational power. In our case, lessening the number of agents reduces computational power but makes it extremely difficult to estimate the field gradient using instantaneous measurements obtained from sensors only. In \cite{Matveev2015}, for example, the authors use a single non-holonomic robot and steer it to an isoline (level curve). The work is gradient-free but the rate of convergence is not satisfactory because the robot ``circulates" along the isoline. The work in \cite{Wu2011} uses 6 agents to elegantly achieve better convergence rates, the latter coming at the cost of using more computational power. The principal attraction of this work is how we fuse these two issues and achieve fast rates of convergence, using a gradient-free control law and minimum computational power.
\par Given the simplicity of our control law, the algorithm has been shown to demonstrate remarkably good results on noisy level curves of relatively complicated functions. The main contribution of the work lies in avoiding the critically heavy step of having to estimate or compute the field gradient. The controller that we develop uses only measurements from the field to track the level curve of a static field in the plane.
\par The rest of the paper is organized as follows. Section II talks about the generalized curve-tracking problem. In Section III, we describe the steps leading to the design of the control law. In Section IV, we perform a stability and convergence analysis for our problem, and derive the conditions when the formation center of our problem converges to the desired level value. Section V presents the results of the algorithm on various level curves and a discussion of the performance of the algorithm. Concluding remarks are presented in Section VI.

\section{Problem Formulation}

Let $z(\mathbf{r})$, $\mathbb{R}^2 \to \mathbb{R}$ denote a scalar field in a two-dimensional space, where $\mathbf{r} \in \mathbb{R}^2$ is the location. Every location of the field is associated with the scalar value of a physical quantity such as temperature or chemical concentration. We have the following assumptions on the field:
\begin{assumption}\label{assumption1}
\begin{enumerate}
\item The field $z(\mathbf{r})$ is smooth with a bounded value, that is, $z_{\mathrm{min}} \leq z(\mathbf{r}) \leq z_{\mathrm{max}}$, where $z_{\mathrm{min}}, z_{\mathrm{max}} \geq 0$.
\item The gradient $\big \Vert \nabla z(\mathbf{r}) \big \Vert \neq 0$ and is bounded, i.e., $\varrho_1\leq\big \Vert \nabla z(\mathbf{r}) \big \Vert \leq \varrho_2$, where $\varrho_1,\varrho_2 > 0$.
%\item The field $z(\mathbf{r})$ is smooth, that is $\dfrac{\partial z}{\partial x}$ and $\dfrac{\partial z}{\partial y}$ both exist, and are both continuous and non-zero.
%\item The norm of the field gradient, $\big \Vert \nabla z(\mathbf{r}) \big \Vert$ is upper bounded by a known upper bound, or that $\big \Vert \nabla z(\mathbf{r}) \big \Vert < \varrho$. It also immediately follows from the positive definiteness property of the norm that $\varrho \geq 0$.
\end{enumerate}
\end{assumption}

%We define a level set $\Omega=\{\mathbf{r}\in\mathbb{R}^2 | z(\mathbf{r})=z_0\}$, in which $z_0$ represents a constant level value. 
Let $\gamma_0(\cdot)$ represent a simple, planar, closed, and regular curve in the field, which is parameterized using its arc length $s$. The length of this curve is finite and is equal to $L$. Then, $s=0$ defines a \textit{starting point} for this curve, which we denote using the point $\mathbf{q}_0(s)$.
%Define $(\mathbf{x}_0(s),\mathbf{y}_0(s))$ to form a right-handed coordinate system with the vector cross product $\Big(\mathbf{y}_0(s) \times \mathbf{x}_0(s)\Big)$ pointing out of the plane of the paper. 
The Frenet-Serret frame \cite{Kuhnel2002} in two dimensions, $(\mathbf{y}_0(s),\mathbf{x}_0(s))$, is traditionally written so as to orient $\mathbf{x}_0(s)$ as the unit tangent vector to the curve and $\mathbf{y}_0(s)$ as the unit normal vector to the curve. Let $\kappa(s)$ be the curvature of the curve such that $\kappa(s_0)$ gives the curvature at $s=s_0$. 
In this setting, the Frenet-Serret equations give the relationship between the frame $(\mathbf{y}_0(s),\mathbf{x}_0(s))$ and the kinematic properties of the curve as
\begin{subequations}\label{Frenet-Serret}
\begin{equation}
\dfrac{\mathrm{d} \mathbf{x}_0(s)}{\mathrm{d}s} = -\kappa(s) \mathbf{y}_0(s),
\end{equation}
\begin{equation}
\dfrac{\mathrm{d} \mathbf{y}_0(s)}{\mathrm{d}s} =  \kappa(s) \mathbf{x}_0(s).
\end{equation}
\end{subequations}
$\gamma_0$ is called a \textit{level curve} of a function $z$ if $z(\gamma_0(\cdot))$ is a constant function of $s$. We assume $\kappa > 0$, which implies that the tangent vector $\mathbf{x}_0$ is moving clockwise.

We consider the problem of estimating the boundary of the field represented by a level curve with a  given level value by deploying two sensing agents in the field. Let $\mathbf{r}_i$ denote the position and $\mathbf{v}_i$ denote the velocity of the $i$th agent. The motion of the agent is constrained by the state dynamics $\dot{\mathbf{r}}_i=\mathbf{v}_i$, $i=1,2$. Suppose these mobile sensing agents are able to take measurements of the field at their current locations, the measurement process being written as
\begin{equation}\label{measurement}
y(\mathbf{r}_i) = z(\mathbf{r}_i) + w(\mathbf{r}_i),
\end{equation}
for $i = 1, 2$. $w(\mathbf{r}_i)$ is assumed to be zero mean Gaussian noise that arises from the measurements or the field itself.  We further assume that each mobile agent has access to the measurements and relative positions of the other agent. The measurements can be exchanged through wireless communication, and the relative locations of other agents can be obtained through cameras, lasers, sonars, etc.

%can exchange information, i.e., the measurements, with all the other mobile agents in the group.

 %has the knowledge of the level value at its own location as well at the location of the other robot.

%\begin{subequations}
%
%\begin{equation}
%\mathbf{r}_c=\frac{1}{N} \sum_{i=1}^N \mathbf{r}_i,
%\end{equation}
%
%\begin{equation}
%\mathbf{v}_c=\frac{1}{N} \sum_{i=1}^N \mathbf{v}_i.
%\end{equation}
%
%\end{subequations}

Denote the formation center of the agents by $\mathbf{r}_c$ and the velocity of the formation center by $\mathbf{v}_c$. We have $\mathbf{r}_c=\frac{1}{2} \sum_{i=1}^2 \mathbf{r}_i$ and $\mathbf{v}_c=\frac{1}{2} \sum_{i=1}^2 \mathbf{v}_i$.

%We would also like to lay down the following assumption about the curvature of the level curves we wish to track:
%\begin{assumption}
%The curvature $\kappa$ of the level curves we are trying to track is upper bounded. In other words, $\exists \: \kappa = \kappa_{\mathrm{max}}$ such that the algorithm is capable of tracking level curves with a curvature less than or equal to $\kappa_{\mathrm{max}}$.
%\end{assumption}

%Consider initially the simple case when $N=2$. In this case, we have two agents which are to be controlled such that they are able to track a level curve in the plane. If the level curve is to pass through the formation center $\mathbf{r}_c$, parameterized using the arc length parameter $s$ (as mentioned above), then $z(\mathbf{r}(s))$ will be a constant for all possible values of $s$. In the setting of the Frenet-Serret equations defined above, we have the unit normal vector $\mathbf{y}_0(s)$ related to the gradient of the curve as
%
%\begin{equation}
%\mathbf{y}_0(s)=\frac{\nabla z(\mathbf{r}(s))}{||\nabla z(\mathbf{r}(s))||}
%\end{equation}
%where the gradient $\nabla z \neq 0$ at any point along the level curve. We also have, $\mathbf{x}_0(s) \cdot \mathbf{y}_0(s)=0$ since the tangent and normal vectors are orthogonal to each other.
%
We define the curve tracking problem using two sensing agents as follows:

{\bf Problem 1:} Consider the motion of the formation center $\mathbf{r}_c$ and the following assumptions:
\begin{itemize}
\item [(A1)] There exists a unique level curve $\gamma_0(s)$ passing through $\mathbf{r}_c$ along the trajectory of $\mathbf{r}_c$.
\item [(A2)] The curvature $\kappa(s)$ of the level curve $\gamma_0(s)$ is bounded at every point of the trajectory of $\mathbf{r}_c$.
\end{itemize}
Given a desired level value $z_d$, design the velocity control of the agents so that the formation center converges to the level curve with value $z_d$ and moves along the curve $\gamma_0(s)$. In other words, design $\mathbf{v}_1$ and $\mathbf{v}_2$, such that $z(\mathbf{r}_c)\to z_d$ as time $t \to \infty$.

We aim to design the control strategy without estimating the field gradient to reduce the computational cost and the sensitivity to noisy measurements. Furthermore, the control strategy should allow the center of the formation to achieve a fast rate of convergence to the level curve. In other words, the formation center converges to a small neighborhood of a desired level curve in finite time, which should be as short as possible.

%Our goal in this work is to design a controller for the velocities of the $N$ agents, $\mathbf{v}_1, \mathbf{v}_2, \cdots, \mathbf{v}_N$, such that this $N-$agent group can successfully track a level curve in the plane. In other words, the final goal of our problem is to design $\mathbf{v}_1, \mathbf{v}_2, \cdots, \mathbf{v}_N$, such that $\frac{\mathbf{v}_c}{||\mathbf{v}_c||} \cdot \frac{\mathbf{x}_0}{||\mathbf{x}_0||} \to 1$ as the time $t \to \infty$ without estimating the field gradient.

\section{Control Law Design}

In this section, we design the control law for the sensing agents so that the formation center tracks a level curve with a desired level value. Assume that associated with each agent $i$, $i=1,2$, we define two unit vectors $\mathbf{q}_i$ and $\mathbf{n}_i$ (perpendicular to $\mathbf{q}_i$) such that $(\mathbf{q}_i,\mathbf{n}_i)$ forms a right-handed coordinate system. The velocities of each individual agent can be decomposed along the $\mathbf{q}_i$ and $\mathbf{n}_i$ directions, so that we can write
\begin{equation}
\mathbf{v}_i=\mathbf{v}_{i,\mathbf{q}}+\mathbf{v}_{i,\mathbf{n}}=v_{i,\mathbf{q}}\mathbf{q}_i+v_{i,\mathbf{n}}\mathbf{n}_i,
\end{equation}
where $v_{i,\mathbf{q}}$ and $v_{i,\mathbf{n}}$ are the projections of $\mathbf{v}_i$ along the directions $\mathbf{q}_i$ and $\mathbf{n}_i$ respectively. We need to design the velocities $v_{i,\mathbf{q}}$ and $v_{i,\mathbf{n}}$, which will successfully guide the center of the two-agent system to track a planar level curve with the desired level value $z_d$.

Since we are considering the case with only two sensing agents, let us define a unit vector $\mathbf{q}$ along the line joining the two agents, i.e., $\mathbf{q} = \frac{\mathbf{r}_2-\mathbf{r}_1}{||\mathbf{r}_2-\mathbf{r}_1||}$, and a unit vector $\mathbf{n}$ being oriented in a way in which $(\mathbf{q},\mathbf{n})$ forms a right-handed coordinate system. Under the assumption that each agent knows the relative position of the other agent, $(\mathbf{q}, \mathbf{n})$ is available to both agents. Therefore, we define $\mathbf{q}_1=\mathbf{q}_2=\mathbf{q}$ and $\mathbf{n}_1=\mathbf{n}_2=\mathbf{n}$.

%we need to consider only the directions $\mathbf{q}$ and $\mathbf{n}$, the former being the unit vector along the line joining the two agents, and the latter unit vector being oriented in a way in which $(\mathbf{q},\mathbf{n})$ forms a right-handed coordinate system. It suffices to consider only the directions $\mathbf{q}$ and $\mathbf{n}$ instead of $\mathbf{q}_1, \mathbf{q}_2, \mathbf{n}_1$ and $\mathbf{n}_2$ because we will resolve the velocities of the two agents along the unique $\mathbf{q}$ and $\mathbf{n}$ direction for the problem.

%We develop the velocity control laws along the $\mathbf{q}$ and $\mathbf{n}$ directions gradually from physical considerations of the problem at hand.

%The $\mathbf{q}$ direction for the problem is given by
%\begin{equation}
%\mathbf{q} = \frac{\mathbf{r}_2-\mathbf{r}_1}{||\mathbf{r}_2-\mathbf{r}_1||}
%\end{equation}
%and the $\mathbf{n}$ direction is given by
%\begin{equation}
%\mathbf{n} = R\mathbf{q}
%\end{equation}
%where $R$ is the rotation matrix $\big( \begin{smallmatrix} 0 & -1 \\ 1 & 0 \end{smallmatrix} \big)$. For a $2-$agent group, we only need the direction vectors $\mathbf{q}$ and $\mathbf{n}$, as these two vectors are enough to resolve the velocities of each agent along two mutually perpendicular directions.

For the sake of notational simplicity, the noisy measurements of the field by the sensors will henceforth be denoted by $y_1$ and $y_2$ instead of $y(\mathbf{r}_1)$ and $y(\mathbf{r}_2)$. $y_1$ and $y_2$ change as the robots move in the plane. Further, we define
\begin{equation} \label{formation_value}
y_c = \frac12 (y_1+y_2).
\end{equation}
%The quantity $y_c$ is brought into the equation for comparison with the desired value $z_d$.
Under Assumption \ref{assumption1}, $y_c$ gives us a satisfactory estimate of the level value of the formation center without having to install a third sensor.

We design the velocity control laws of the two mobile robots along the $\mathbf{q}$ direction as
\begin{equation}\label{eqn:vq}
v_{i,\mathbf{q}} = k_1((\mathbf{r}_j - \mathbf{r}_i) \cdot \mathbf{q} - d_{i,j}^0) + k_2 \mathrm{sgn}((y_c - z_d)(y_1 - y_2)),
\end{equation}
for $i = 1,2$, $j = 1,2$, $i \neq j$, where $k_1 > 0$ is a constant, $d_{i,j}^0=-d_{j,i}^0=d^0$ is the desired distance between the two agents, and
%and $u_1$ is modeled as
%\begin{equation}\label{eqn:vq_u}
%u_1 = k_2 \mathrm{sgn}((y_c - z_d)(y_1 - y_2)).
%\end{equation}
 the function $\mathrm{sgn}(\cdot)$ is defined as
\[
    \mathrm{sgn}(x) = \left\{\begin{array}{lr}
        -1, & \text{if } x < 0\\
        0, & \text{if } x = 0\\
        1, & \text{if } x > 0
        \end{array} \right.
  \]
where the constant $k_2 > 0$.

%\begin{subequations}
%
%\begin{equation}
%v_{1,\mathbf{q}} = k_1((\mathbf{r}_2 - \mathbf{r}_1) \cdot \mathbf{q} - d_0) + u_1 ,
%\end{equation}
%
%\begin{equation}
%v_{2,\mathbf{q}} = -k_1((\mathbf{r}_2 - \mathbf{r}_1) \cdot \mathbf{q} - d_0) + u_1 .
%\end{equation}
%
%\end{subequations}
The first term in Equation \eqref{eqn:vq} is for formation control, that is to ensure that the mobile robots maintain a distance of $d_0$ between each other along the $\mathbf{q}$ direction at equilibrium. The constant $k_1 > 0$ determines the rate of convergence of the agents. The second term aims to enable the agents to track the level curve with value $z_d$ in the plane.
\begin{figure}[!t]
\centering
\includegraphics[width=0.2\textwidth]{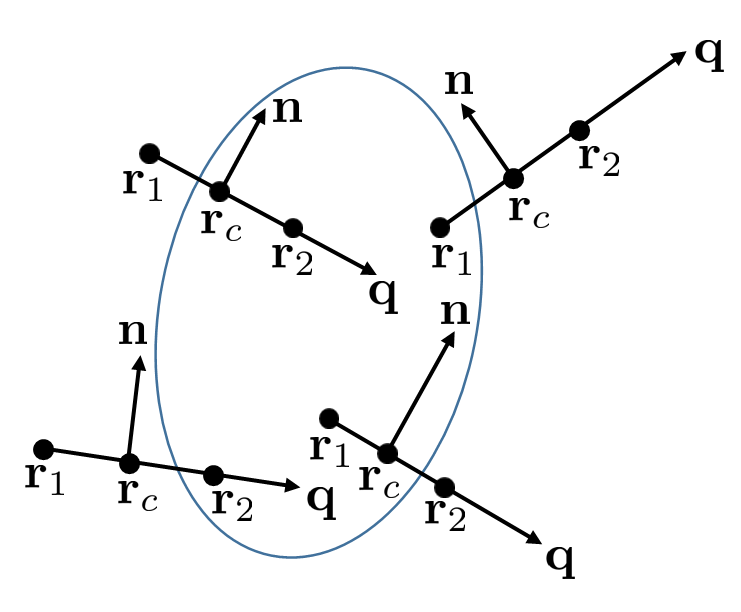}
\caption{Positions assumed by the two-robot system when tracking an ellipse in the plane.}
\label{fig:formation}
\end{figure}
The motivation behind the modeling of the second term can be justified from Fig. \ref{fig:formation}. The movement of the agents along the $\mathbf{q}$ direction depends on the relative positions of the agents, and the relative position of the formation center with respect to the level curve.
%Since the motion is dependent on both these factors, we use the product of signum functions to guide the system along the $\mathbf{q}$ or $-\mathbf{q}$ direction.
Accordingly, we find, for example in the case in the top left, $\mathrm{sgn}(y_1 - y_2) = 1$ and $\mathrm{sgn}(y_c - z_d) = -1$. So the formation moves along the $-\mathbf{q}$ direction, as required. The veracity of the way in which the second term has been modeled can be checked from the remaining three cases illustrated in Fig. \ref{fig:formation} as well.
% Here, $z_d$ is the level value which we seek to track. As we already know, each mobile robot has the knowledge of the level value at its own location as well as at the location of the other robot.

Next, we design the velocity control law along the $\mathbf{n}$ direction as
%What remains is the velocity control law along the $\mathbf{n}$ direction. We present the control law first and then argue for the reason behind its form. We have

\begin{equation}\label{eqn:vn}
v_{i,\mathbf{n}} = \begin{cases} 
   C + ay_i, & \text{if } \big| y_c - z_d \big| < \varepsilon \\
   0,      & \text{otherwise}
  \end{cases}
\end{equation}
%\begin{equation}
%v_{i,\mathbf{n}} = \left\{\begin{array}{lr}
%        C + ay_i, & \text{if } \big| y_c - z_d \big| < \varepsilon \\
%        0, & \text{otherwise } \\
%        \end{array} \right.
%\end{equation}
where $i = 1,2$, $j = 1,2$, $i \neq j$, $C > 0$ and $a > 0$. After we've achieved formation control using the two-robot system and we're `away' from the level curve in some capacity, we'd want the two-agent system to move along the $\mathbf{q}$ direction only, to first come within an $\varepsilon-$distance of the level curve. Once the latter objective is achieved, we would want forward motion of the agents along the level curve. When the formation center is `close' enough to the level curve $-$ which is ensured by the condition $\big| y_c - z_d \big| < \varepsilon$ $-$  we impart velocities to the agents along the $\mathbf{n}$ direction. Consider again the case on the upper left hand side in Fig. \ref{fig:formation}. Since $y_1 > y_2$, we'll have $v_{1,\mathbf{n}} > v_{2,\mathbf{n}}$. This will result in a net clockwise torque, which will guide the system along the curve.

%The reason behind modeling the $u_1$ term the way it is modeled can be justified from the figure presented above. We present (without loss of generality) four cases which may arise when the two-robot system is tracking an ellipse on the plane. We've already stated that we assume here that each agent has the knowledge of the field value which the other agent knows. Consequently, we need to control the velocity along the `$\mathbf{q}$' direction and push the mobile robot system either along $\mathbf{q}$ or along $-\mathbf{q}$ depending on the level curve values sampled by the two robots. Consider the case shown on the top left-hand side of Figure \ref{Formation_Demo}. Assuming that the level values, $z$, increase from the center outwards, we have $z_1 > z_2$ and $z_c < z_d$. This means that $\mathrm{sgn}(z_1-z_2) = 1$ and $\mathrm{sgn}(z_c - z_d) = -1$. Under these cases, the control law pushes the mobile robot system towards the $-\mathbf{q}$ direction, which is indeed what is needed for the formation center to stay on the level curve. We leave the reader to verify the veracity of the control law for the other cases illustrated.

\section{Stability and Convergence Analysis}

We will perform our analysis assuming the complete absence of noisy measurements or corrupted field values. In other words, we assume that $w(\mathbf{r}_i) \equiv 0$ in Equation (\ref{measurement}). Therefore, in the following analysis, we replace $y_i, i=1,2$ and $y_c$ in the control law \eqref{eqn:vq} and \eqref{eqn:vn} with $z_i, i=1,2$ and $z_c$, respectively. We will verify the performance of our controller in the presence of corrupted field values in the simulation. 

The presence of the signum functions in Equation (\ref{eqn:vq}) as well as the modeling of the $v_{i,\mathbf{n}}$ term in Equation (\ref{eqn:vn}) make our control law discontinuous. We can look at our control law in some capacity as a sliding mode controller, in which a discontinuous control law makes the system slide along a sliding manifold. Accordingly, we have two cases, one for which the formation center is away from the level curve, and the other for which the formation center is close to the level curve. 

Based on the control law, when the formation center is away from the desired level curve, i.e., $|z_c-z_d|\geq\epsilon$, the forward speed of the agents $v_{i,\mathbf{n}}=0$. Therefore, $\mathbf{v}_i=v_{i,\mathbf{q}}\mathbf{q}$, which leads to 
\begin{equation}
\mathbf{v}_c=\frac{1}{2}(\mathbf{v}_1+\mathbf{v}_2)=k_2 \mathrm{sgn}((z_c - z_d)(z_1 - z_2))\mathbf{q},
\end{equation}
since the first term in \eqref{eqn:vq} cancels. 
\begin{figure}[!t]
\centering
\includegraphics[width=0.2\textwidth]{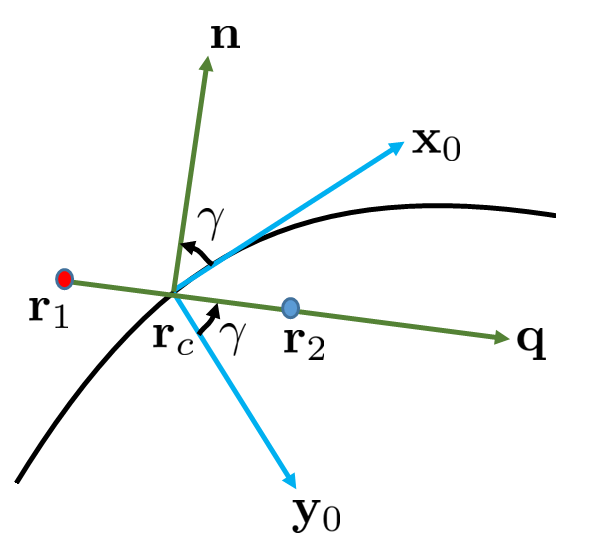}
\caption{Orientations of the coordinate frames $(\mathbf{q},\mathbf{n})$ and $(\mathbf{y}_0,\mathbf{x}_0)$. Rotating the latter frame by an angle $\gamma$ gives the former frame. The level curve is shown in black. The two agents are denoted by the red and blue circles.}
\label{fig:coordinate}
\end{figure}
Once the formation center reaches the neighborhood of a given level curve, i.e., $|z_c-z_d|<\epsilon$, the forward speed of the agents is proportional to the field value, i.e., $v_{i,\mathbf{n}}=C+az_i$. Therefore, the velocity of the formation center becomes
\begin{equation}
\mathbf{v}_c = \Big( k_2 \mathrm{sgn}((z_1 - z_2)(z_c - z_d)) \Big) \mathbf{q} + \Big(C+az_c \Big) \mathbf{n}.
\end{equation}
where, analogous to Equation \eqref{formation_value}, we have $z_c = \tfrac{1}{2}(z_1+z_2)$. Denote $\alpha \triangleq k_2 \mathrm{sgn}((z_1 - z_2)(z_c - z_d))$ and $\beta \triangleq C + az_c$, then $\mathbf{v}_c$ can be written as
\begin{equation}
\mathbf{v}_c = \alpha \mathbf{q} + \beta \mathbf{n}.
\end{equation}

We will show that under the designed control law, the formation center will converge to the level curve with the desired level value $z_d$ by properly choosing the values $k_2$, $C$, and $a$. In this paper, we consider the case when the direction of $\nabla z_c$ is anti-parallel to that of $\mathbf{y}_0$ and the two-agent formation is moving in a clockwise direction along a level curve, as illustrated in Fig. \ref{fig:coordinate}. This setting implies that $z_1>z_2$ if the two-agent formation is moving along a level curve. The analysis we will show can be immediately extended to other settings, e.g., when the two-agent formation moves counterclockwise and $\nabla z_c$ is parallel to that of $\mathbf{y}_0$, after proper changes of the signs of the notations.

We first derive the dynamic equation for the field value at the center of the two-agent formation. We have
\begin{equation}\label{eqn:dz_c}
\dot{z}_c = \frac{\partial z_c}{\partial \mathbf{r}_c} \cdot \dot{\mathbf{r}}_c = \nabla z_c \cdot \dot{\mathbf{r}}_c = \nabla z_c \cdot \mathbf{v}_c = \nabla z_c \cdot (\alpha \mathbf{q} + \beta \mathbf{n}).
\end{equation}
Writing $\nabla z_c = -\big\Vert \nabla z_c \big\Vert \mathbf{y}_0,$
leads to
\begin{equation}\label{zc_state}
\dot{z}_c = -\big\Vert \nabla z_c \big\Vert \mathbf{y}_0 \cdot (\alpha \mathbf{q} + \beta \mathbf{n}) = -\big\Vert \nabla z_c \big\Vert (\alpha (\mathbf{y}_0\cdot\mathbf{q}) + \beta (\mathbf{y}_0\cdot{\mathbf{n}})).
\end{equation}

Note from Figure \ref{fig:coordinate} that the $(\mathbf{q},\mathbf{n})$ frame is obtained from the $(\mathbf{y}_0,\mathbf{x}_0)$ frame by a rotation of $\gamma$. In order to show the convergence of $z_c$, we further derive the dynamic equation for $\mathbf{n}\cdot\mathbf{x}_0$. We have
\begin{equation}\label{eqn:dnx0}
{\mathrm{d}(\mathbf{n}\cdot\mathbf{x}_0) \over \mathrm{d}t} = \dot{\mathbf{n}} \cdot \mathbf{x}_0 + \mathbf{n} \cdot \dot{\mathbf{x}}_0.
\end{equation}
On the one hand, $\dot{\mathbf{n}}$ and $\dot{\mathbf{q}}$ can be found out from the Frenet-Serret relations for the frame $(\mathbf{q},\mathbf{n})$
\begin{subequations}\label{Frenet-Serret-1}
\begin{equation}
\dot{\mathbf{n}} = \omega \mathbf{q},
\end{equation}
\begin{equation}
\dot{\mathbf{q}} = -\omega \mathbf{n},
\end{equation}
\end{subequations}
where $\omega$ is the angular velocity of the formation center. 
Denoting $\sigma=\frac{\mathrm{d}s}{\mathrm{d}t}$ as the speed of the formation center, we have
\begin{subequations}\label{Frenet-Serret-2}
\begin{equation}
\dot{\mathbf{x}}_0 =-\sigma \kappa\mathbf{y}_0,
\end{equation}
\begin{equation}
\dot{\mathbf{y}}_0 = \sigma\kappa \mathbf{x}_0.
\end{equation}
\end{subequations}
$\sigma$ is related to $\alpha$ and $\beta$ by $\sigma=\sqrt{\alpha^2+\beta^2}$. 

To find the angular velocity $\omega$, we express $z_1$ and $z_2$ using Taylor expansion with respect to the center $\mathbf{r}_{c}$, that is, $z_i=z_c+\nabla z_c\cdot(\mathbf{r}_i-\mathbf{r}_c)+O(\|\mathbf{r}_i-\mathbf{r}_c\|^2)$, in which $O(\cdot)$ represents ``on the order of". 
Note that $\|\mathbf{r}_2-\mathbf{r}_1\|=2\|\mathbf{r}_2-\mathbf{r}_c\|=2\|\mathbf{r}_1-\mathbf{r}_c\|$.
Then, the angular velocity can be approximated as
\begin{align}\label{eqn:omega}
\omega&=\frac{v_{2,\mathbf{n}}-v_{1,\mathbf{n}}}{\|\mathbf{r}_2-\mathbf{r}_1\|}=\frac{a(z_2-z_1)}{\|\mathbf{r}_2-\mathbf{r}_1\|} \nonumber\\
&=\frac{a\nabla z_c\cdot(\mathbf{r}_2-\mathbf{r}_1)+O(\|\mathbf{r}_2-\mathbf{r}_1\|^2)}{\|\mathbf{r}_2-\mathbf{r}_1\|}\nonumber\\
&=-a\|\nabla z_c\|(\mathbf{y}_0\cdot\mathbf{q})+O(\|\mathbf{r}_2-\mathbf{r}_1\|).
\end{align}
Note that since we are able to select the two points $\mathbf{r}_2$ and $\mathbf{r}_1$ to be arbitrarily close to the center, the term $O(\|\mathbf{r}_2-\mathbf{r}_1\|)$ can be made arbitrarily small.
Therefore we will omit this term in the rest of the analysis.
Combining Equations \eqref{Frenet-Serret-1}, \eqref{Frenet-Serret-2}, and \eqref{eqn:omega}, Equation \eqref{eqn:dnx0} becomes 
\begin{equation}\label{eqn:dnx0_1}
{\mathrm{d}(\mathbf{n}\cdot\mathbf{x}_0) \over \mathrm{d}t} = -a\|\nabla z_c\| (\mathbf{y}_0\cdot\mathbf{q})(\mathbf{q} \cdot \mathbf{x}_0) - \kappa \sigma (\mathbf{n} \cdot \mathbf{y}_0).
\end{equation}

Since $\mathbf{n} \cdot \mathbf{x}_0 = \cos\gamma$, $\mathbf{n} \cdot \mathbf{y}_0 = -\sin\gamma$, $\mathbf{q} \cdot \mathbf{x}_0 = \sin\gamma$ and $\mathbf{q} \cdot \mathbf{y}_0 = \cos\gamma$, 
%So, $\mathbf{n} \cdot \mathbf{x}_0 = \theta$, $\mathbf{n} \cdot \mathbf{y}_0 = \cos (\tfrac{\pi}{2}+\gamma) = - \sqrt{1-\theta^2}$, $\mathbf{q} \cdot \mathbf{x}_0 = \cos (\tfrac{\pi}{2}-\gamma) = \sqrt{1-\theta^2}$ and $\mathbf{q} \cdot \mathbf{y}_0 = \theta$. 
$\omega$ can be approximated by
%\begin{equation}
%\phi = (\alpha \mathbf{q} + \beta \mathbf{n}) \cdot \mathbf{x}_0 = \alpha \sqrt{1-\theta^2} + \beta \theta
%\end{equation}
%and,
\begin{equation}
\omega \approx -a\|\nabla z_c\|\cos\gamma,
\end{equation}
and the dynamic equations \eqref{eqn:dz_c} and \eqref{eqn:dnx0_1} can be rewritten as 
\begin{equation}\label{eqn:zc_state}
\dot{z}_c =  -\big\Vert \nabla z_c \big\Vert (\alpha\cos\gamma- \beta \sin\gamma),
\end{equation}
and
\begin{equation}\label{eqn:cosgamma}
{\mathrm{d}\cos\gamma \over \mathrm{d}t} =-a\|\nabla z_c\|\cos\gamma\sin\gamma+\kappa\sigma\sin\gamma.
%&=- (\pm a\|\nabla z_c\|\theta\sqrt{1-\theta^2}\mp \kappa\sigma\sqrt{1-\theta^2}).
\end{equation}
We have the following lemma regarding the angle $\gamma$. 

\begin{lemma}\label{lemma_gamma}
Suppose at $t=0$, we set $\gamma(0) \in \left( 0, \tfrac{\pi}{2} \right)$.
% then we can ensure that $\gamma(t) \in \left[ -\tfrac{\pi}{2}, \tfrac{\pi}{2} \right] \: \forall \: t$ as $t \to \infty$.
There exists $b\in(0,1)$ such that if at $t=0$, $\cos\gamma\in(0,b)$, then $\cos\gamma$ will rise to be greater than $b$ in finite time, i.e., $\cos\gamma>b$, and stay in $(b,1]$ after that, if, at any time instant, the speed of the formation center satisfies $\sigma>\frac{a\|\nabla z_c\|b}{\kappa(\mathbf{r}_c)}$, where $\kappa(\mathbf{r}_c)$ is the curvature of the level curve passing through the formation center. Correspondingly, $\gamma$ will stay in $[0,\arccos b)$. 
\end{lemma}
\begin{IEEEproof}
Consider the dynamic equation for $\cos\gamma$ in \eqref{eqn:cosgamma}. When $\cos\gamma=b$, we have $\sin\gamma=\sqrt{1-b^2}$, which corresponds to $\gamma=\arccos b \in(0,\frac{\pi}{2})$, Equation \eqref{eqn:cosgamma} becomes 
\begin{equation}\frac{\mathrm{d}\cos\gamma}{\mathrm{d}t} =(-a\|\nabla z_c\|b+\kappa\sigma)\sqrt{1-b^2}>0\end{equation} under the assumption that $\sigma>\frac{a\|\nabla z_c\|b}{\kappa(\mathbf{r}_c)}$. Therefore, $\cos\gamma$ will rise above $b$ in finite time. When $\cos\gamma=1$, Equation \eqref{eqn:cosgamma} becomes 
$\frac{\mathrm{d}\cos\gamma}{\mathrm{d}t} = 0$, which means that $\gamma$ stops changing. Therefore, $\cos\gamma$ will stay in $(b,1]$ afterwards. 
\end{IEEEproof}

We are now ready to state and prove our main result:

\begin{theorem}\label{main_theorem}
Define the closed (metric) annulus with $\varepsilon > 0$ centered at a point $z_d$ in a set $M$, $\mathcal{A}_{\varepsilon}[z_d] = \{ z_c \in M \:\:\: | \:\:\: z_d - \varepsilon \leq d(z_c,z_d) \leq z_d + \varepsilon \}$ where the metric space $(M,d)$ is any set $M$ equipped with the ordinary Euclidean distance function $d$. If at $t=0$, we set $\gamma(0) \in \left( 0, \tfrac{\pi}{2} \right)$, then the center of the formation will converge to the level curve with the level value $z_d$ asymptotically from the boundaries of the annulus $\mathcal{A}_{\varepsilon}[z_d]$, if the choices of $C$, $a$, and $k_2$ in the controller \eqref{eqn:vq} and \eqref{eqn:vn} and the field value at the formation center satisfies $C+az_{\mathrm{max}}<~\frac{k_2b}{\sqrt{1-b^2}}$.
\end{theorem}
\begin{IEEEproof}
Consider the Lyapunov candidate function
\begin{equation}
V = \frac12 (z_c-z_d)^2.
\end{equation}
$V(z_c=z_d)=0$ and $V>0$ for $z_c \neq z_d$.
We then have,
\begin{align}
\dot{V} &= (z_c - z_d) \dot{z}_c, \nonumber
\\ &= - (z_c - z_d) \| \nabla z_c \| (\alpha \cos \gamma - \beta \sin \gamma).
\end{align}
Since we are considering the case that the two-agent system is moving clockwise along a level curve in a smooth field, $z_1-z_2>0$. Therefore, the sign of $\alpha$ will be determined by $z_c-z_d$. We investigate the sign of $\dot{V}$ at the boundaries of $\mathcal{A}_{\varepsilon}[z_d]$, namely at $z_c = z_d + \varepsilon$ and $z_c = z_d - \varepsilon$. When $z_c = z_d + \varepsilon$, $\alpha=k_2$ since $z_c - z_d = \varepsilon > 0$. From Lemma \ref{lemma_gamma}, there exists $b$ such that $\cos\gamma>b$ and $\sin\gamma<\sqrt{1-b^2}$.
Therefore, 
\begin{align}
\dot{V} &= - \varepsilon \| \nabla z_c \| (k_2 \cos \gamma - \beta \sin \gamma), \nonumber\\
&< - \varepsilon \| \nabla z_c \| (k_2 b - \beta\sqrt{1-b^2}). 
\end{align}
Plug $\beta=C+az_c$ into $\dot{V}$. Under the condition that $C+az_{\mathrm{max}}<\frac{k_2b}{\sqrt{1-b^2}}$, we obtain 
\begin{equation}
\dot{V} < - \varepsilon \| \nabla z_c \| (k_2 b - (C+az_c)\sqrt{1-b^2})<0. 
\end{equation}
When $z_c = z_d - \varepsilon$, $\alpha=-k_2$ since $z_c - z_d = -\varepsilon < 0$. In this case, 
\begin{align}
\dot{V} &= - \varepsilon \| \nabla z_c \| (-k_2 \cos \gamma - \beta \sin \gamma), \nonumber\\
&= - \varepsilon \| \nabla z_c \| (k_2 \cos \gamma + \beta \sin \gamma)<0.
\end{align}
So, we have that $\dot{V}\Big|_{z_c = z_d + \varepsilon} < 0$ and $\dot{V}\Big|_{z_c = z_d - \varepsilon} < 0$. Hence, when the formation center converges towards the desired level value $z_d$ from the boundaries of the annulus $\mathcal{A}_{\varepsilon}[z_d]$.
\end{IEEEproof}
We therefore have the results that under the conditions of Lemma \ref{lemma_gamma} and Theorem \ref{main_theorem}, the formation center asymptotically converges to the desired level value $z_d$. The reason behind proving Lemma \ref{lemma_gamma} is because having bounds on the $\cos \gamma$ and $\sin \gamma$ terms helps us to efficiently handle these trigonometric terms that arise from the state equation \eqref{eqn:zc_state} associated with $z_c$.

\par It is also interesting to note that although we may intuitively expect from Fig. \ref{fig:coordinate} that the vector $\mathbf{n}$ will have the same direction as the vector tangent to the level curve $\mathbf{x}_0$ as $t \to \infty$ (or that $\gamma$ asymptotically converges to $0$ and $\cos \gamma$ asymptotically converges to $1$), in reality this is not true. The reason is as follows. Substituting $\frac{\mathrm{d} \cos \gamma}{\mathrm{d} t}$ with $ -\sin \gamma \dot{\gamma}$ into Equation \eqref{eqn:cosgamma}, we have:
%\[\frac{\mathrm{d} \cos \gamma}{\mathrm{d} t} = -a \| \nabla z_c \| \cos \gamma \sin \gamma + \kappa \sigma \sin \gamma,\]
%\begin{align}
%&\implies -\sin \gamma \dot{\gamma} = -a \| \nabla z_c \| \cos \gamma \sin \gamma + \kappa \sigma \sin \gamma, \nonumber
%\\ &\implies \dot{\gamma} = a \| \nabla z_c \| \cos \gamma - \kappa \sigma.
%\end{align}
\begin{equation}
-\sin \gamma \dot{\gamma} = -a \| \nabla z_c \| \cos \gamma \sin \gamma + \kappa \sigma \sin \gamma,
\end{equation}
which leads to
\begin{equation}
\dot{\gamma} = a \| \nabla z_c \| \cos \gamma - \kappa \sigma,
\end{equation}
if $\sin\gamma\neq0$.
Letting $\dot{\gamma} = 0$ gives us
\begin{equation}
\gamma = \arccos \left( \frac{\kappa(t) \sigma(t)}{a \| \nabla z_c(t) \|} \right),
\end{equation}
where the explicit dependence on time is shown to indicate that the value of $\gamma$ changes at every instant of time. Therefore, the angle between the frames $(\mathbf{q},\mathbf{n})$ and $(\mathbf{y}_0,\mathbf{x}_0)$ does not converge to a time-invariant value. We will demonstrate experimental validation of this in the next section.

\section{Simulation Results}

The performance of the algorithm is tested on two functions, one of them being an ellipse, and the other one being a function which has a relatively more complicated landscape, and, for this reason, is used in testing the performance of optimization algorithms.

\par
\begin{figure}[!t]
\centering
\includegraphics[width=0.44\textwidth]{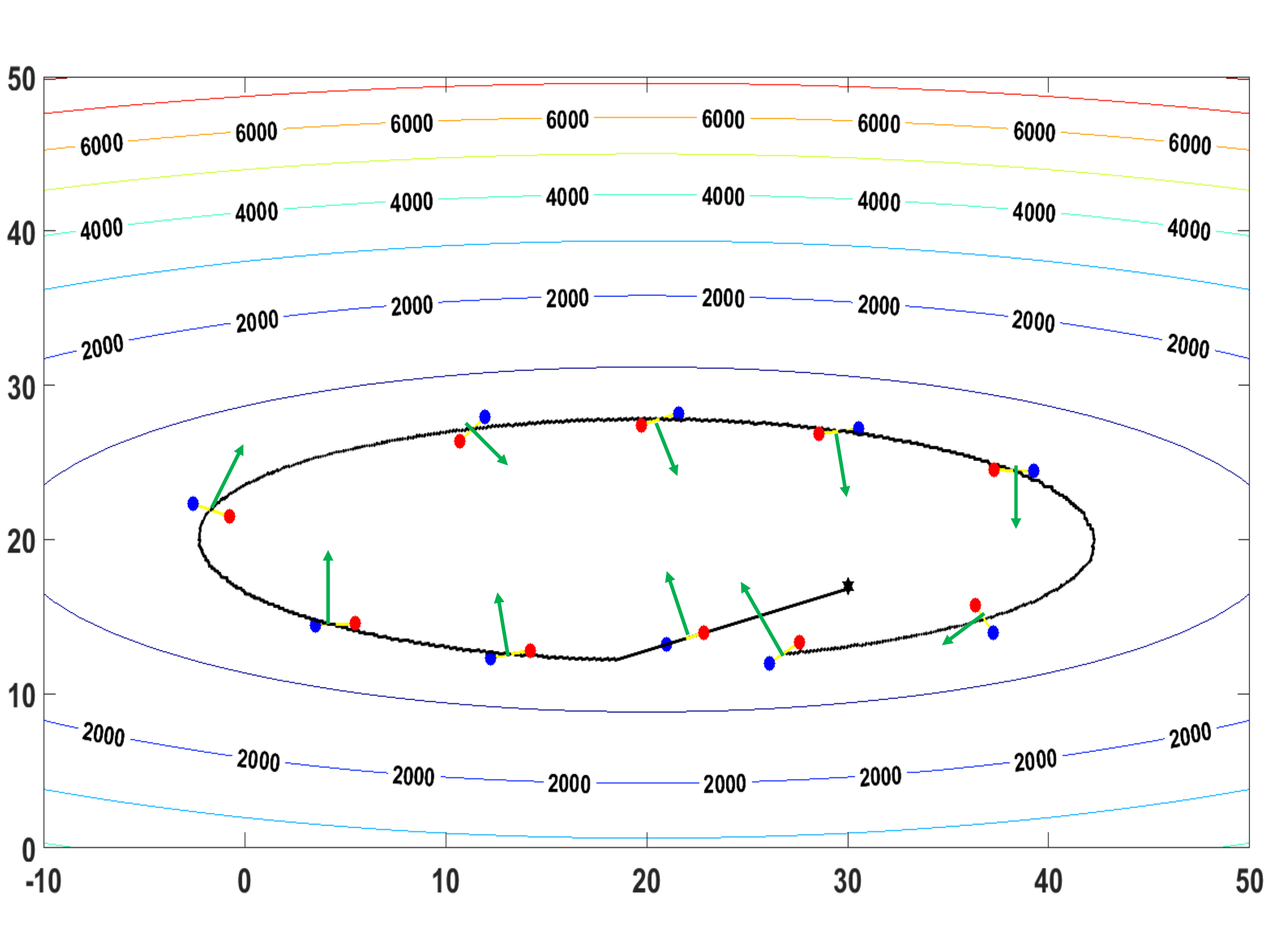}
\caption{Trajectories of a 2-agent group tracking $z_d = 500$ for the ellipse $z = (x-20)^2 + 8(y-20)^2$. }
%The directions of the green arrows (but not their magnitude) are indicative of the direction of the vector $\mathbf{n}$.}
\label{Traj_Ellipse}
\end{figure}

\begin{figure}[!t]
\centering
\includegraphics[width=0.33\textwidth]{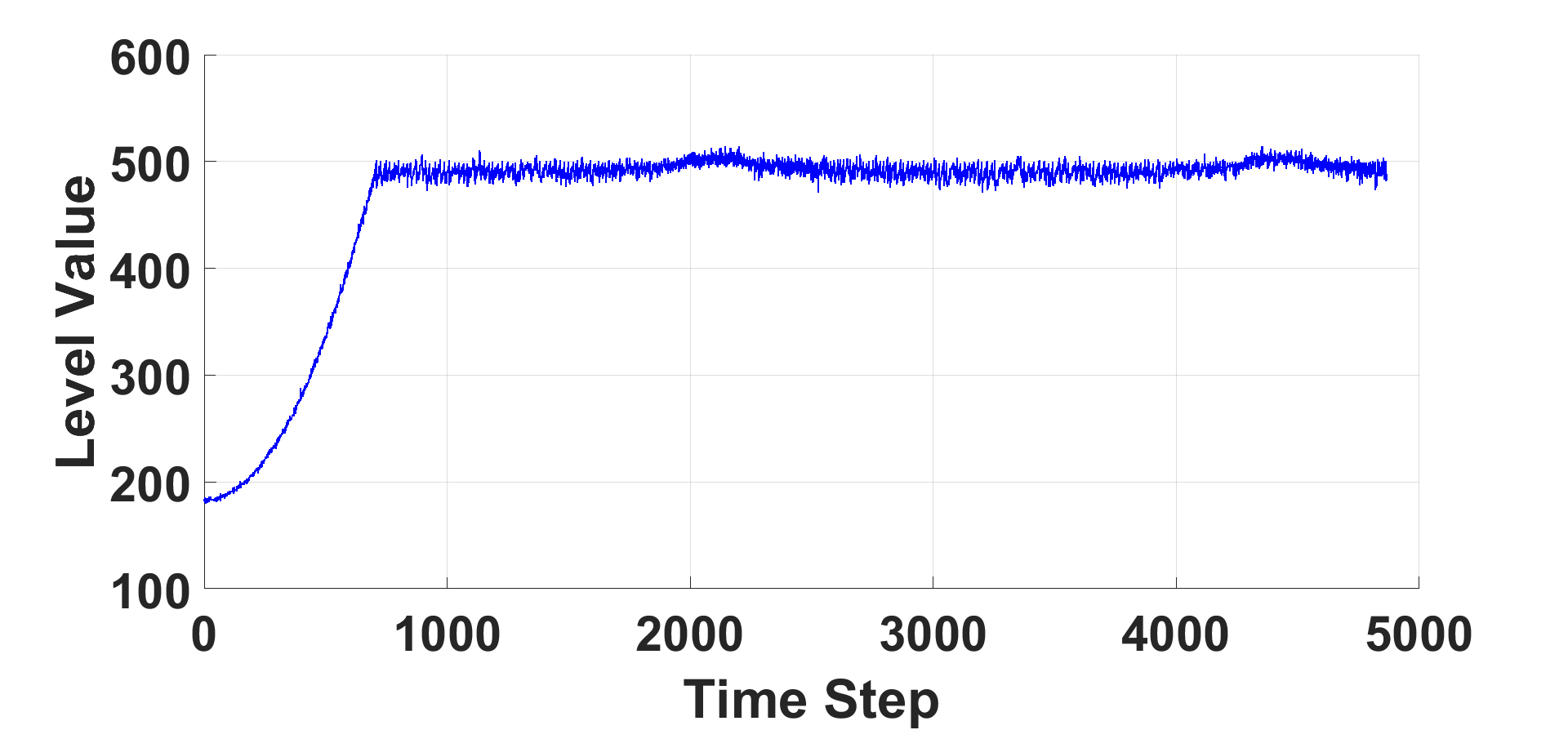}
\caption{Evolution of tracked level value at the formation center with time step for tracking $z_d = 500$ for the ellipse $z = (x-20)^2 + 8(y-20)^2$.}
\label{Evol_Ellipse}
\end{figure}

 Figs. \ref{Traj_Ellipse} and \ref{Evol_Ellipse} show the performance of the algorithm on the level curves of the ellipse $z = (x-20)^2 + 8(y-20)^2$. We track the level curves having a level value of 500. In Fig. \ref{Traj_Ellipse}, the red and blue dots represent the two mobile sensing agents and the trajectory of the formation center is shown in black.  Noise has been incorporated into the simulation by adding normally distributed random numbers to each field value. We use the values $k_1 = 1$, $k_2 = 0.9$, $C = 1$, $a = 0.01$ and $\varepsilon = 2$. The two-agent system converges to the desired level value $z_d = 500$ quickly and smoothly, and the formation center is able to track the desired level curve to a high degree of accuracy while staying in the vicinity of the curve. The green arrows indicate the direction of the vector $\mathbf{n}$. In Fig. \ref{Evol_Ellipse}, the evolution of the level value at the formation center is tracked as a function of time.

\par
\begin{figure}[!t]
\centering
\includegraphics[width=0.44\textwidth]{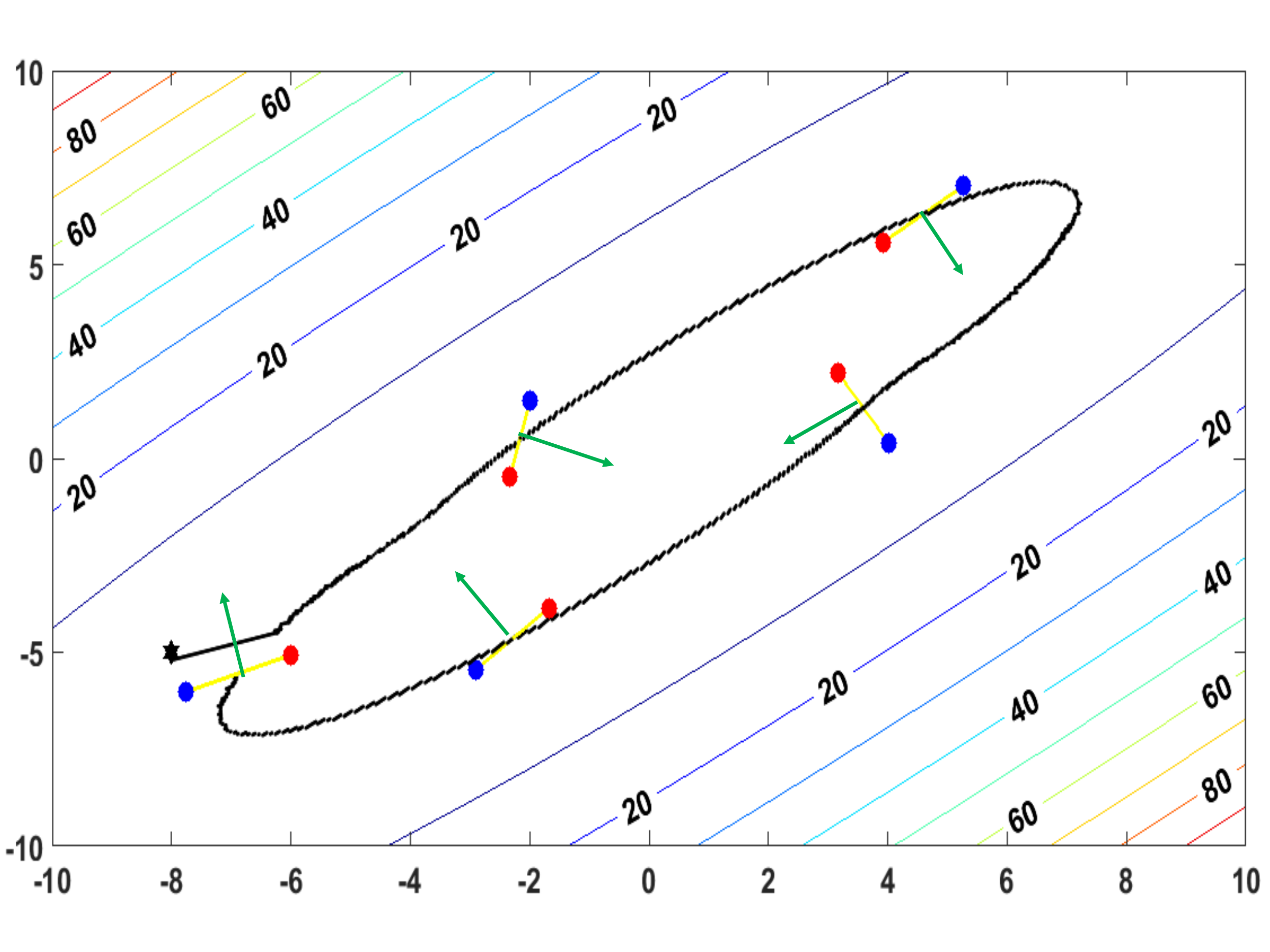}
\caption{Trajectories of a 2-agent group tracking $z_d = 2$ for the Matyas function $z = 0.26(x^2 + y^2) - 0.48xy$. }
%The directions of the green arrows (but not their magnitude) are indicative of the direction of the vector $\mathbf{n}$.}
\label{Traj_Matyas}
\end{figure}

\begin{figure}[!t]
\centering
\includegraphics[width=0.3\textwidth]{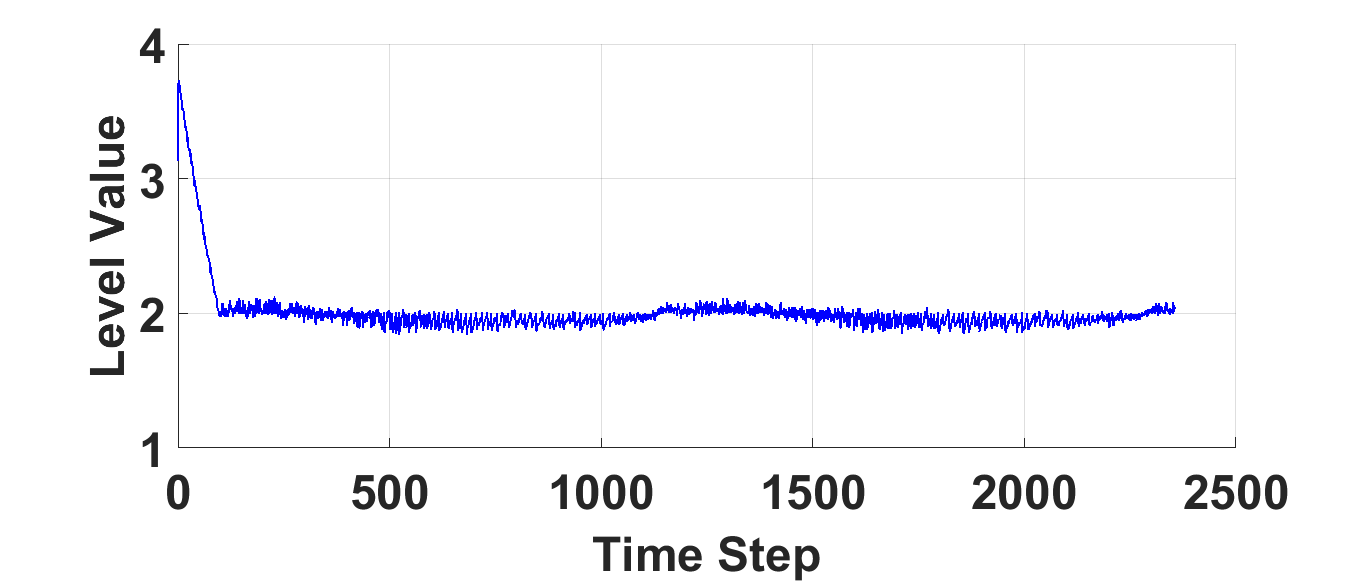}
\caption{Evolution of tracked level value at the formation center with time step for tracking $z_d = 2$ for the Matyas function $z = 0.26(x^2 + y^2) - 0.48xy$.}
\label{Evol_Matyas}
\end{figure}

 The control law is also tested on the Matyas Function, which is specified by the relation $z = 0.26(x^2 + y^2) - 0.48xy$. Figs. \ref{Traj_Matyas} and \ref{Evol_Matyas} show the performance of our algorithm on level curves of the Matyas function having a level value of 2. We use $k_1 = 1$, $k_2 = 0.99$, $C = 1$, $a = 1$ and $\varepsilon = 0.01$ as values of the constants for this simulation, and also add noise in the form of normally distributed random values to the field value. Fig. \ref{Evol_Matyas} shows the evolution of the level value at the formation center as time increases. The algorithm demonstrates convergence in sufficiently small time even for relatively complicated landscapes. Here too, the direction of the vector $\mathbf{n}$ is shown by the green arrows.

\par We note from the directions of the vector $\mathbf{n}$ in the Figs. \ref{Traj_Ellipse} and \ref{Traj_Matyas} that as the formation center moves along the level curve, the vector $\mathbf{n}$ indeed does not converge to $\mathbf{x}_0$ as $t \to \infty$. We thus obtain experimental confirmation of the fact that the angle between the frames $(\mathbf{q},\mathbf{n})$ and $(\mathbf{y}_0,\mathbf{x}_0)$ does not converge to a time-invariant equilibrium value.
\par Hence, we see from the above examples that our control law is able to demonstrate a good performance on level curves of many types of two-dimensional functions. From the last example, we find that the algorithm performs remarkably well even while tracking level curves of functions having non-trivial complicated landscapes. The motivation behind the development of the control law is the absence of a term where the field gradient has to be estimated since that proves to be the most computationally expensive step in these kinds of problems. In addition, the work deals with solving the problem using two agents only, and hence using minimum computational power.

%\begin{figure}[!t]
%\centering
%\includegraphics[width=0.55\textwidth]{McCormick}
%\caption{Tracking the McCormick Function $z = \sin(x + y) + (x - y)^2 - 1.5x + 2.5y + 1$: Tracking the level curve.}
%\end{figure}

\section{Conclusion}
In this work, we develop an algorithm to successfully track planar level curves using two agents without explicitly estimating the field gradient. Based on physical considerations, we decompose the velocity of each agent into two mutually perpendicular directions, and then develop the velocity control law along each direction. The velocity control is developed in a way that the two-agent group can successfully track a level curve in the plane without having to explicitly estimate the gradient of the scalar field in question. Our results show that even in the case of level curves of relatively complicated functions, the two-agent group is able to track the level curve with a high degree of accuracy and a fast rate of convergence towards the level curve. Possible future directions include extending the work to one where we tackle the general problem of using $N$ agents to track a level curve without explicitly computing the field gradient.


%\section*{Acknowledgment}
%The acknowledgment goes here...
%



%\bibliographystyle{IEEEtran}

\bibliographystyle{IEEEtran}
\bibliography{IEEEabrv,Bibliography}

\end{document}